\documentclass[twocolumn,preprintnumbers,amsmath,amssymb]{revtex4}

\usepackage{graphicx}
\usepackage{dcolumn}
\usepackage{bm}

\begin{document}

\title{Collective Dynamics of Active Elements: \\
Task Allocation and Pheromone Trailing}

\author{T. Mizuguchi}
\affiliation{Department of Mathematical Sciences, Osaka Prefecture University}%
\author{K. Sugawara}%
\affiliation{Department of Liberal Arts, Tohoku Gakuin University}%
\author{H. Nishimori}%
\affiliation{Department of Mathematical Sciences, Osaka Prefecture University}%
\author{T. Tao}%
\affiliation{Department of Mathematical Sciences, Osaka Prefecture University}%
\author{T. Kazama}%
\affiliation{Graduate School of Information Systems, University of Electro-Communications}%
\author{H. Nakagawa}%
\affiliation{Department of Mathematical Sciences, Osaka Prefecture University}%
\author{Y. Hayakawa}%
\affiliation{Department of Physics, Tohoku University}%
\author{M. Sano}%
\affiliation{Department of Physics, University of Tokyo}%

\date{\today}

\begin{abstract}
Collective behavior of active elements 
inspired by mass of biological organisms is addressed. 
Especially, two topics are focused on among 
amazing behaviors performed by colony of ants. 
First, task allocation phenomena are treated from the viewpoint 
of proportion regulation of population between different states. 
Using a dynamical model consisting of elements 
and external ``stock materials'', 
adaptability against various disturbances is numerically studied. 
In addition,
a dynamical model for a colony ants interacting via two kind of pheromones is 
studied, in which simulated ants, as a mass, are shown to make an efficient foraging 
flexibly varying the foraging tactics according to feeding schedules.  
Finally, 
experiments are performed with robots moving in 
virtual pheromone fields simulated by CG and CCD camera feedback system. 
Trail formation processes are demonstrated by this 
multi-robot system.
\end{abstract}

\maketitle

\section{Introduction}
Among intriguing phenomena exhibited by biological mass, 
it is remarkable that various functions or 
performances of the whole mass 
exceed simple superposition of each individual's ability. 
The mass of individuals we consider here is that of 
slime molds, (social) insects, fish, birds, etc. 
For example, a kind of cellular slime mold shows 
regulation of ratio between two different types of cell
during its differentiation process 
\cite{Raper,Bonner,Loomis,Oyama}.
About the motion of the group, 
birds flock, fish school and mosquitoes swarm
\cite{Wilson2, Edelstein-Keshet, Partridge, Inoue, Vicsek, Shimoyama}.
Social insects exhibit lots of amazing performances 
like colony formation, age polyethism, group foraging, etc\cite{Wilson}. 
It is worth noticing that these collective performances 
do not require special individual(s) 
which controls the behavior of the whole. 
Hereditarily homogeneous individuals achieve 
these collective behaviors 
by interacting  each other through direct sight 
or chemical materials like c-AMP, pheromones, etc. 
How does each of them ``feel'' the situation of the whole and 
how does it determine its behavior?
Here, we focus on the task allocation phenomena 
and the pheromone trail formation processes 
among various collective behaviors described above. 

This article is organized as follows. 
In the next section, 
from the viewpoint of proportion regulation, 
multi-task allocation phenomena are treated 
with a dynamical model consisting of simple 
elements and external variables. 
In \S 3, foraging  behavior of modeled ants was treated in which 
ants leave and follow only the local  information of pheromone, however 
they can fulfill the highly efficient foraging 
flexibly adapting to various feeding conditions.
In \S 4, real robots experiments are performed. 
Using a virtual pheromone system (V-DEAR) which consists of 
LC projector and CCD camera, 
both task allocation process and trail formation process 
are examined by multi-robot system.
Summary is given in \S 5.

\section{Task allocation}

Division of labor ({\it polyethism}) performed by social insects 
or cell differentiation observed in the life cycle 
of {\it Dictyostelium discoideum} exhibit 
typical proportion regulation phenomena.
Namely, individuals (or cells) with the same inheritance 
divide into two or more groups and they play different roles. 
The proportion of populations of each group is regulated 
against external disturbance without special individual. 
Regulation means a kind of adaptability, 
i.e., for large disturbance like loss of some part of colony 
or change environment, 
individuals know the change of the situation 
about the whole and change their behavior. 
As a result of such autonomous process, 
the proportion of population between the groups is maintained. 
Several questions naturally arise for these phenomena, i.e., 
how do they know about the whole and how do they determine their behavior?
In this section, dynamical models for internal states of individuals and 
external variables of chemical materials 
are used to reproduce these phenomena 
and consider their mechanism through numerical simulations. 

As a previous work, we suggest a dynamical model for 
proportion regulation phenomena for two states. 
The model consists of $N$ identical elements with 
two internal variables, 
i.e., activator $u_i(t)$ and inhibitor $v_i(t)$, 
couple globally through their average quantities\cite{GlobalTuring}.
The dynamics of the internal variables $u_i(t)$ and $v_i(t)$ is
\begin{eqnarray}
\dot{u_i} &=& au_i - bv_i - u_i^3 + K_1(\bar{u} - u_i), \\
\dot{v_i} &=& cu_i - dv_i         + K_2(\bar{v} - v_i), 
\end{eqnarray}
where $\bar{\phantom{u}}$ denotes averaging over population, 
i.e. , $\bar{u}\equiv \sum u_j / N$ 
and $\bar{v}\equiv \sum v_j/N$. 
$a,b,c,$ and $d$ are non-negative parameters. 
$K_1$ and $K_2$ are the non-negative susceptibilities 
of each variables in the context of fast diffusion limit.
The model has following properties under suitable 
parameter conditions $a-d<0$, $ad-bc<0$ and $K_1=0$: 
(i) The population ``differentiates'' into two states for $K_2>K_{2c}$. 
(ii) The population ratio between the two states 
is regulated within some range which is obtained analytically. 
(iii) The system maintains the ratio against large 
disturbance by changing the state of individuals.
(iv) State of each element depends only on its initial condition. 
The model, however, is only available for two states allocation 
because of its simplicity. 
In order to treat multi-states case, 
we adopt another dynamical model in the next subsection.

\subsection{Model}

\begin{figure}[t]
\includegraphics[width=8cm]{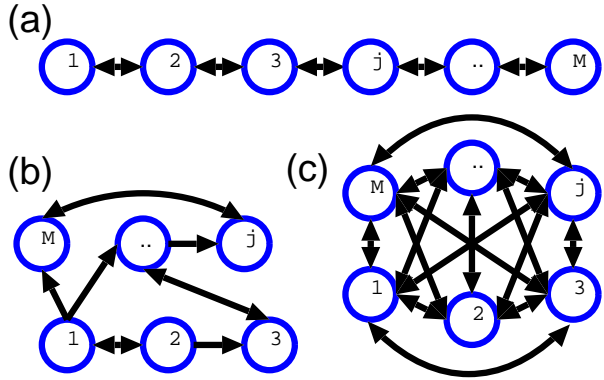}
\caption{\label{fig:rules}
Examples of rule between multi-states. 
(a) sequential (b) intermediate, and (c) democratic.
Arrows denote the path which the rule allows to change. 
Note that the paths between the states in rule (a) and (c) 
are all symmetric while it is not in (b). 
}
\end{figure}

\begin{figure}[h]
\begin{picture}(7,5)(0,0)
\put(0.,0.0){\includegraphics[width=7cm]{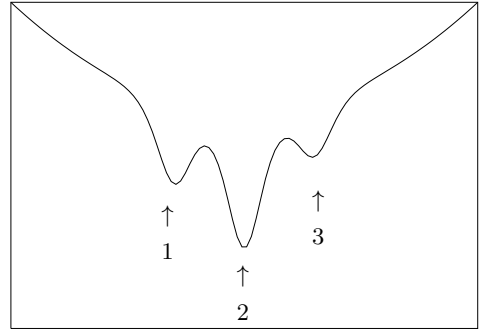}}
\put(2.4,1.7){$\uparrow$}\put(2.4,1.2){1}
\put(3.4,0.9){$\uparrow$}\put(3.4,0.4){2}
\put(4.4,1.9){$\uparrow$}\put(4.4,1.4){3}
\end{picture}
\caption{\label{fig:potential}
Typical shape of potential function $U$ with three states.
}
\end{figure}

Unlike the two states case, 
multi-states case requires detail setting up the problem, 
i.e., not only the ratio between the states 
but also the rule between the states are needed to be assigned. 
Here, the rule is considered to be 
determined by the connectivity between the states. 
Several examples of rule are shown in Fig.\ref{fig:rules}. 
We assume the sequential and symmetric rule (Fig.\ref{fig:rules}(a)) here, 
and set up the situation as follows: 
$N$ identical individuals, 
$M$-states with proportion of population $p_j$ for $j$-th state
with $0 < p_j < 1$ and $\sum p_j = 1$.
$M=3$ and $p_j=\{0.5, 0.3, 0.2\}$ are used as typical setting up.

Our model consists of $N$ identical elements and $M$ ``stock materials''. 
The elements represent the individuals and 
have internal variable $u_i(t)$, $i=1,...,N$. 
The stock materials correspond the states and 
their quantities are expressed by external variable $w_j(t)$ , $i=1,...,M$. 
Each state is expressed by a valley of potential function $U(u)$, 
\begin{eqnarray}
U(u) &=& \sum_j (\alpha_j-w_j(t)) U_j(u) + \alpha_0 U_0(u),\\
U_j(u) &=& -{e^{-\beta(u-s_j)} \over(1+e^{-\beta(u-s_j)})^2},\\
U_0(u) &=& u^2, 
\end{eqnarray}
where $s_j$ and $\alpha_j$ denote the positions and depths 
of $j$-th valley, respectively (Fig.\ref{fig:potential}). 
$\beta$ gives an average depth of the valleys. 
These quantities are determined by the environment. 
$U_0$ is used for boundedness of the internal variables 
of individuals.
$w_j(t)$, the quantity of $j$-th stock materials ,
changes effective depth of $j$-th potential and 
obeys following equation 
\begin{eqnarray}
\tau_j \dot{w}_j &=& -w_j + a \sum_i(-U_j(u_i)), 
\end{eqnarray}
where the first term of the R.H.S represents decay process 
and the second term is production process by the individuals 
staying in the corresponding state. 
$\tau_j$ is characteristic time of $j$-th material
and $a$ is a production rate. 
The dynamics of $u_i(t)$, the internal variable of $i$-th element, is 
\begin{eqnarray}
\dot{u}_i &=& - {\partial U(u) \over \partial u} + \eta,
\end{eqnarray}
where $\eta$ is a white uniform noise with amplitude $\nu$. 
Individuals are considered to be subjected under the 
time dependent potential $U(u)$ and the noise $\eta$.

\begin{figure}[h]
\begin{picture}(8,12.5)(0,0)
\put(1.3,  9.2){\frame{\includegraphics[height=2.5cm,width=4.9cm]{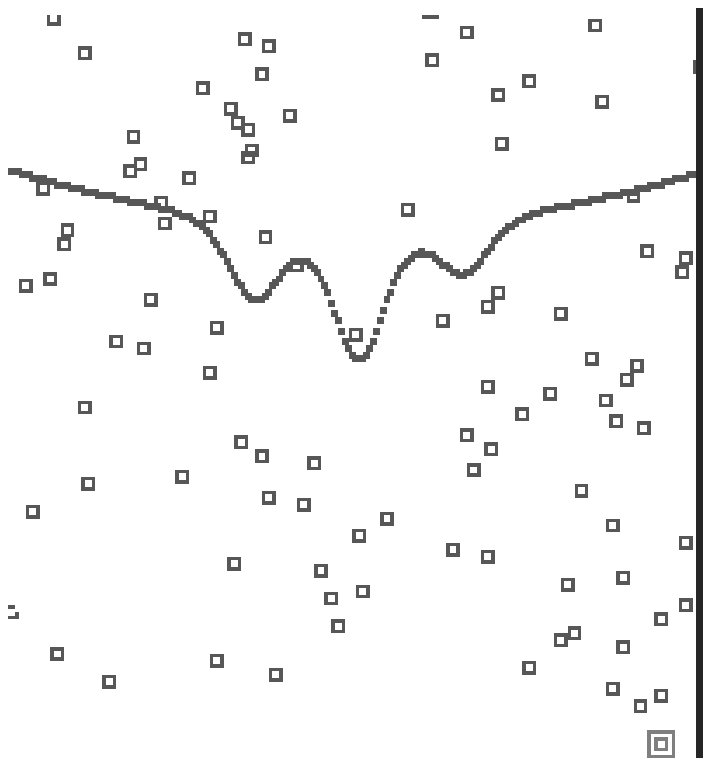}}}
\put(1.3,  6.5){\frame{\includegraphics[height=2.5cm, width=4.9cm]{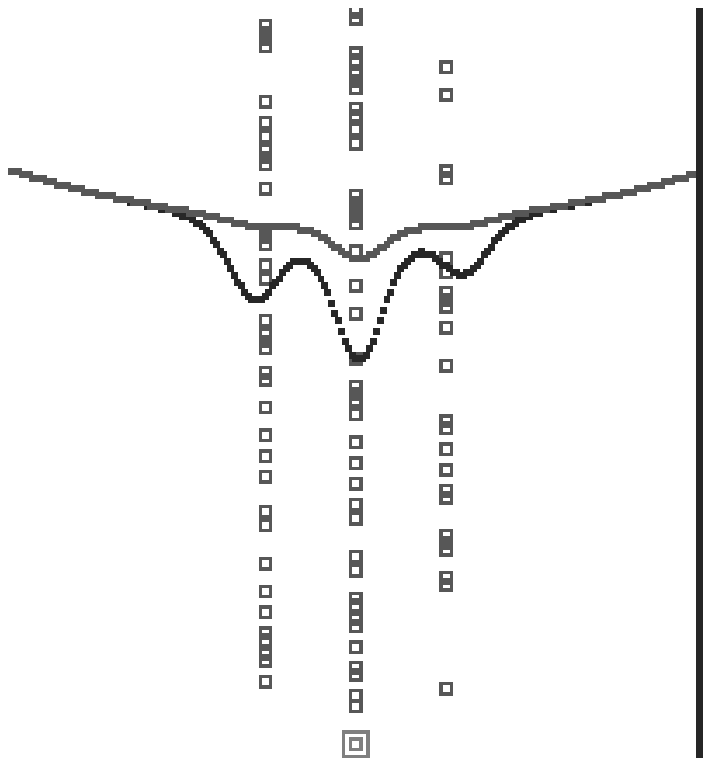}}}
\put(1.3, 11.3){\normalsize (a)}
\put(1.3,  8.7){\normalsize (b)}
\put(3.5,  6.1){\normalsize $u_j$}
\put(1.1,  6.1){\normalsize $-1$}
\put(5.9,  6.1){\normalsize $1$}
\put(0.8, 11.2){\normalsize $j$}
\put(6.3, 11.2){\normalsize $U$}
\put(0.8,  8.6){\normalsize $j$}
\put(6.3,  8.6){\normalsize $U$}
\put(5.0, 11.3){\normalsize $t=0$}
\put(4.6,  8.7){\normalsize $t=100$}
\put(1.2,  0.5){\rotatebox{90}{\includegraphics[width=5cm]{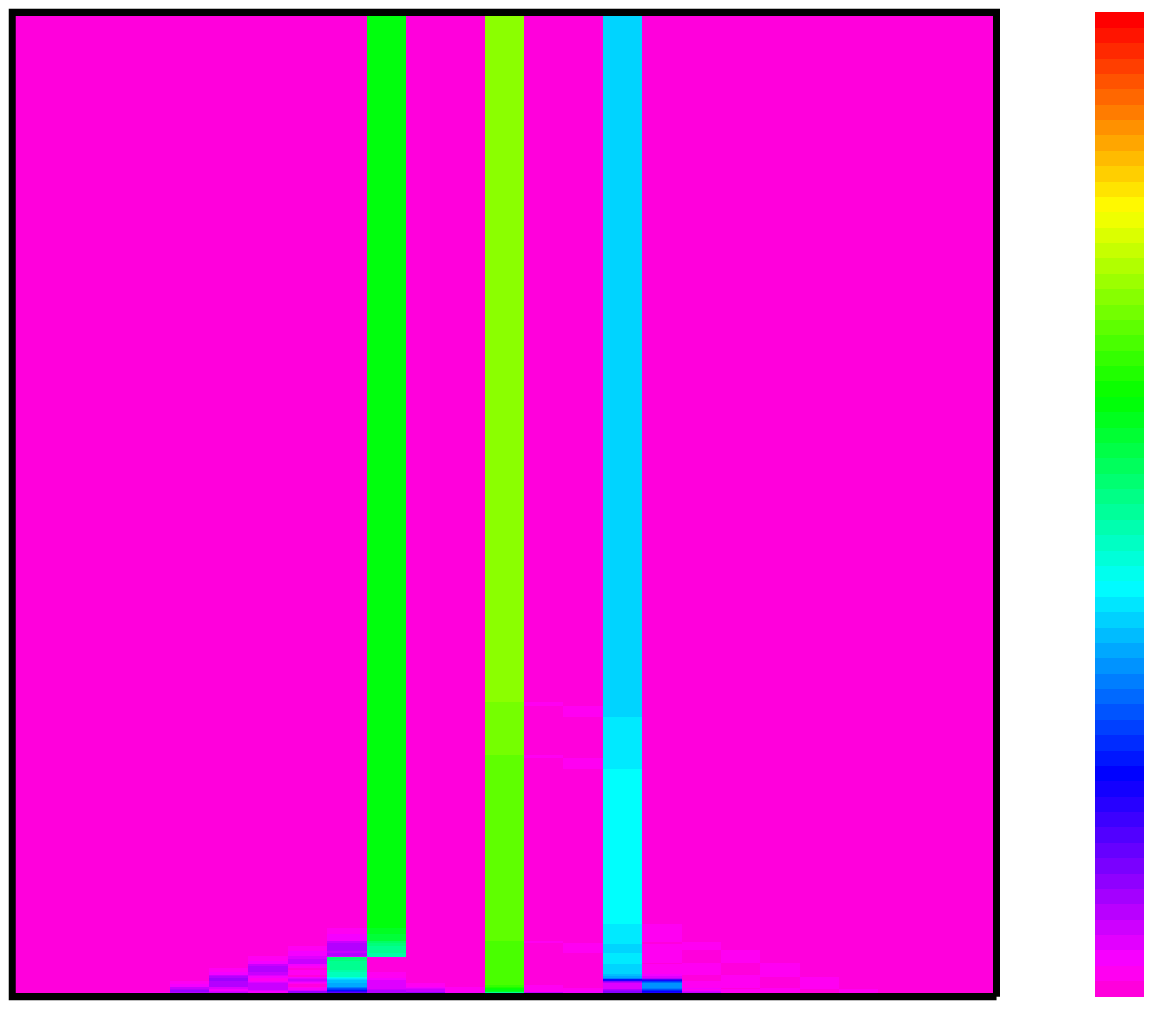}}}
\put(1.3, 5.1){\normalsize (c)}
\put(3.6, 0.1){\normalsize $u_j$}
\put(1.2, 0.1){\normalsize $-1$}
\put(6.0, 0.1){\normalsize $1$}
\put(0.6,2.8){\normalsize $t$}
\put(0.9,0.5){\normalsize $0$}
\put(0.5,5.3){\normalsize $100$}
\put(7.0, 0.5){\normalsize $0$}
\put(7.0, 5.3){\normalsize $60$}
\end{picture}
\caption{\label{fig:noiseless}
State of $j$-th element $u_j$ (open rectangle) and 
the shape of the potential at $t=0$ (a) and $t=100$ (b). 
The potential is modified by stock materials $w_j$. 
To display the modification by $w_j$, 
initial shape of the potential ($w_j=0$) are also plotted in (b).
(c) Evolution of distribution function. 
Color scale denotes number of individuals from $0$ to $60$.
}
\end{figure}

\subsection{Noiseless case}

We now show the results of numerical simulations. 
Following conditions are used in this and the next subsections:
$N=100$, $a=3.5$, $\beta=20.0$, $\tau_j=1$, 
$\alpha_0=0.1$, $\alpha_1=0.3$, $\alpha_2=0.5$, $\alpha_3=0.2$, 
$s_1=-0.3$, $s_2=0$ and $s_3=0.3$. 
Typical snapshots and the evolution of the system 
are displayed in Fig.\ref{fig:noiseless} for the noiseless case $(\nu = 0)$. 
As an initial condition, 
the internal variable $u_j(0)$ of each element is randomly distributed 
and the stock materials $w_j(0) $ are all set be zero (Fig.\ref{fig:noiseless}(a)). 
Each element feels the potential $U(t)$ 
and gathers toward the valley for each state. 
Simultaneously, the elements which staying at the $j$-th valley 
product the corresponding stock materials $w_j$ and 
change effective depth of the potential. 
Designed ratio (3:5:2) is approximately achieved and 
the shape of the potential becomes almost flat 
at $t=100$ (Fig.\ref{fig:noiseless}(b)). 
The evolution of distribution function shows  
the designed ratio is almost established 
in early stage (Fig.\ref{fig:noiseless}(c)).

\begin{figure}
\begin{picture}(8,15)(0,0)
\put(1.5,10.5){\rotatebox{90}{\includegraphics[height=6cm,width=4cm]{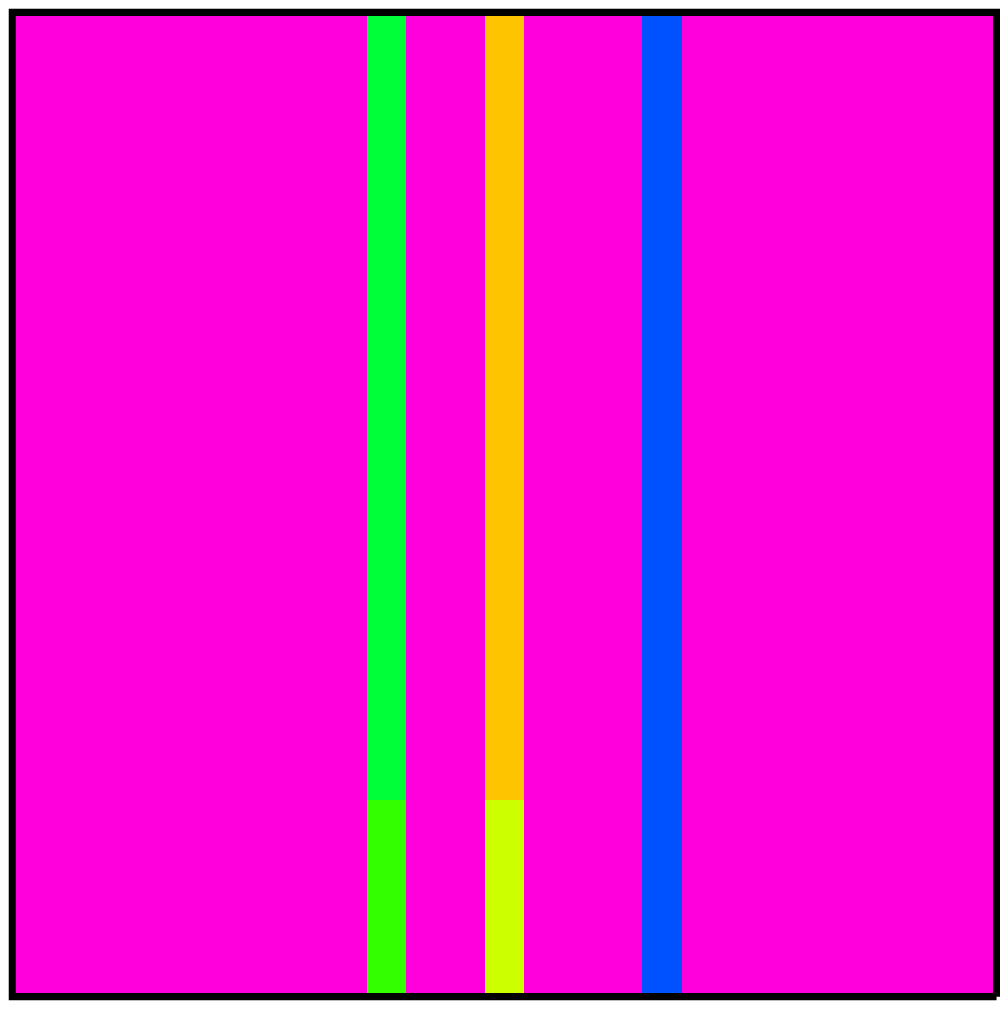}}}
\put(1.7,14.0){\normalsize (a)}
\put(4.1,10.1){\normalsize $u_j$}
\put(1.4,10.1){\normalsize $-1$}
\put(6.6,10.1){\normalsize $1$}
\put(0.9,12.2){\normalsize $t$}
\put(1.1,10.5){\normalsize $0$}
\put(0.8,14.2){\normalsize $100$}
\put(1.5,11.2){\normalsize $\rightarrow$}
\put(1.1,11.2){\normalsize $t^*$}
\put(1.5,5.5){\rotatebox{90}{\includegraphics[height=6cm,width=4cm]{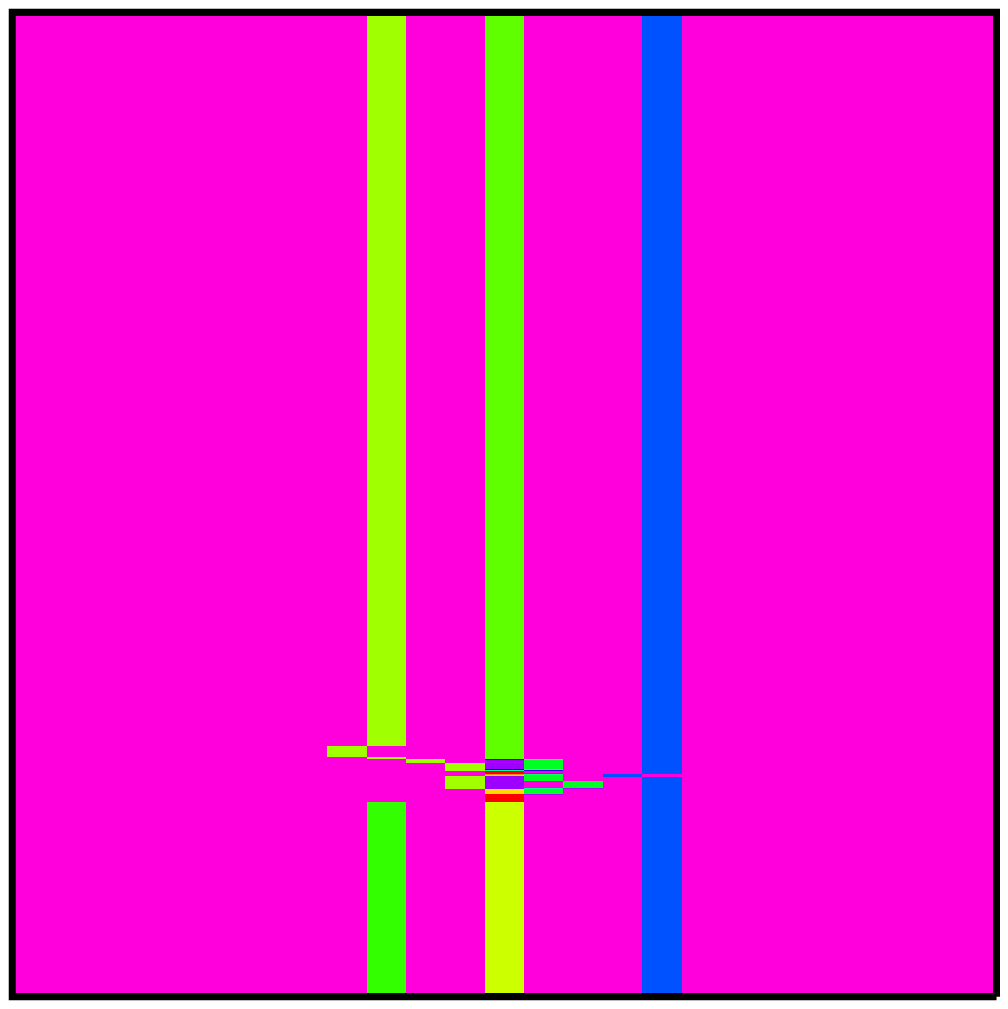}}}
\put(1.7,9.0){\normalsize (b)}
\put(4.1,5.1){\normalsize $u_j$}
\put(1.4,5.1){\normalsize $-1$}
\put(6.6,5.1){\normalsize $1$}
\put(0.9,7.2){\normalsize $t$}
\put(1.1,5.5){\normalsize $0$}
\put(0.8,9.2){\normalsize $100$}
\put(1.5,6.2){\normalsize $\rightarrow$}
\put(1.1,6.2){\normalsize $t^*$}
\put(1.5,0.5){\rotatebox{90}{\includegraphics[height=6cm,width=4cm]{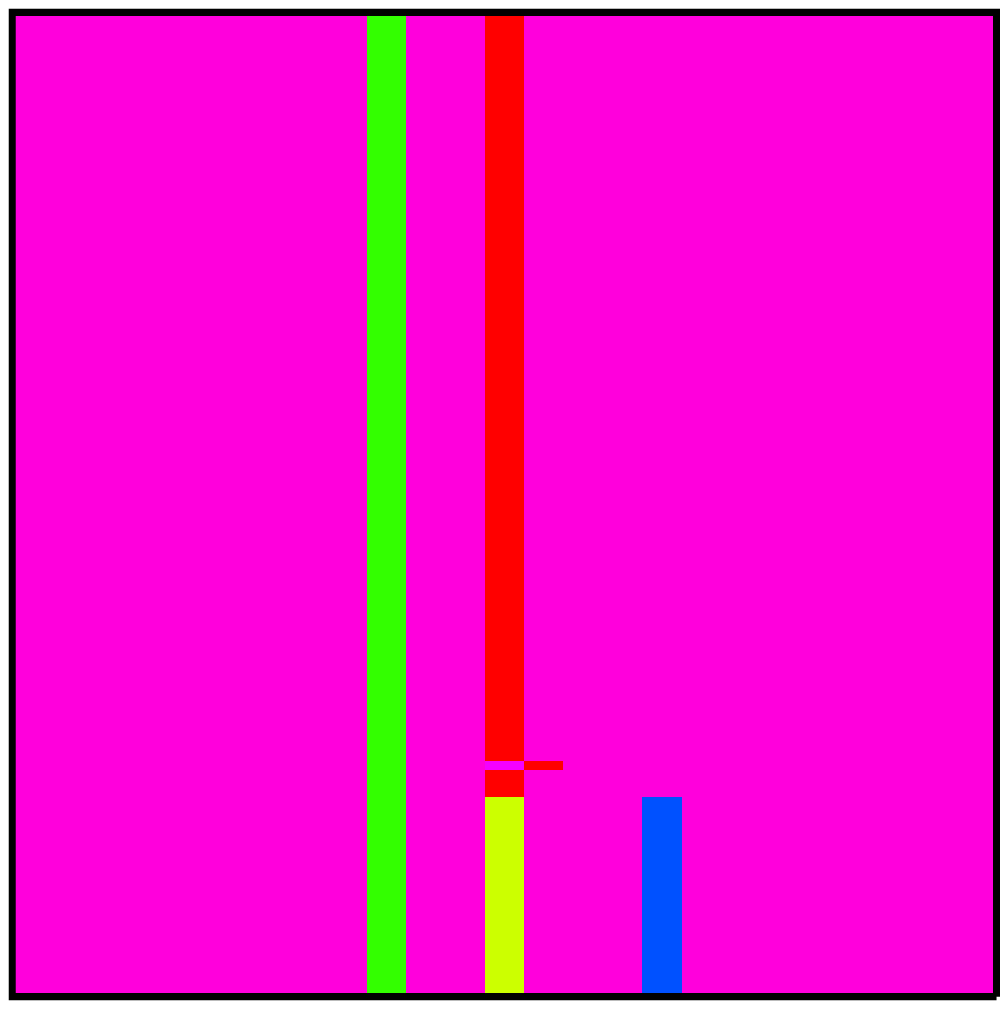}}}
\put(1.7,4.0){\normalsize (c)}
\put(4.1,0.1){\normalsize $u_j$}
\put(1.4,0.1){\normalsize $-1$}
\put(6.6,0.1){\normalsize $1$}
\put(0.9,2.2){\normalsize $t$}
\put(1.1,0.5){\normalsize $0$}
\put(0.8,4.2){\normalsize $100$}
\put(1.5,1.2){\normalsize $\rightarrow$}
\put(1.1,1.2){\normalsize $t^*$}
\end{picture}
\caption{\label{fig:noiseless2}
Response for the disturbances added at $t^*$. 
The evolution of distribution function is displayed. 
(a) Insensitivity for small disturbances. 
(b) Adaptability for large disturbances. 
(c) Invalid case against large disturbances. 
}
\end{figure}

The response of the system for various type of 
disturbance is shown in Fig.\ref{fig:noiseless2}. 
First, a few elements in state 3 are moved to 
state 2 at $t^*$ as a small disturbance. 
The system is insensitive and reconstruction 
of distribution does not occur (Fig.\ref{fig:noiseless2}(a)). 
Next, large disturbance is given, i.e., 
all the elements in state 1 are moved to state 2. 
In this case, the reconstruction process occurs and 
the ratio between the states are regulated again 
(Fig.\ref{fig:noiseless2}(b)). 
Finally, an inadequate example is presented. 
When all the elements in state 3 are moved to state 2, 
undesired two clusters is realized (Fig.\ref{fig:noiseless2}(c)).
This is because the system is trapped in local minimum of the potential. 
This fault can be avoided by the noise term as shown in the next subsection. 

\begin{figure}
\begin{picture}(6,9)(0,0)
\put(0.5,5){\rotatebox{90}{\includegraphics[height=5.5cm,width=4cm]{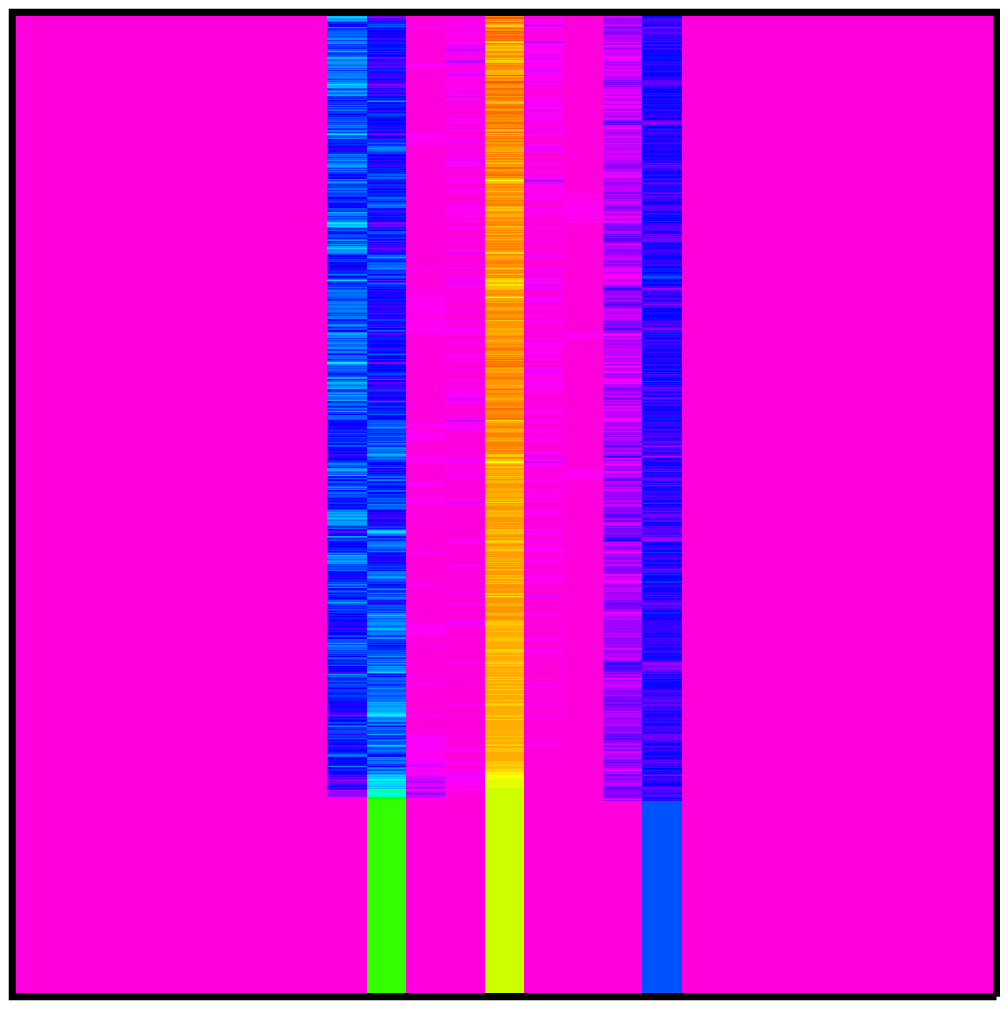}}}
\put(0.7,8.5){\normalsize (a)}
\put(2.9,4.6){\normalsize $u_j$}
\put(0.4,4.6){\normalsize $-1$}
\put(5.2,4.6){\normalsize $1$} \put(-0.1,6.9){\normalsize $t$}
\put(0.1,5.0){\normalsize $0$}
\put(-0.2,8.7){\normalsize $100$}
\put(0.5,5.7){\normalsize $\rightarrow$}
\put(0.1,5.7){\normalsize $t^*$}
\put(0.2,0){\reflectbox{\rotatebox{90}{\includegraphics[height=5.4cm,width=4cm]{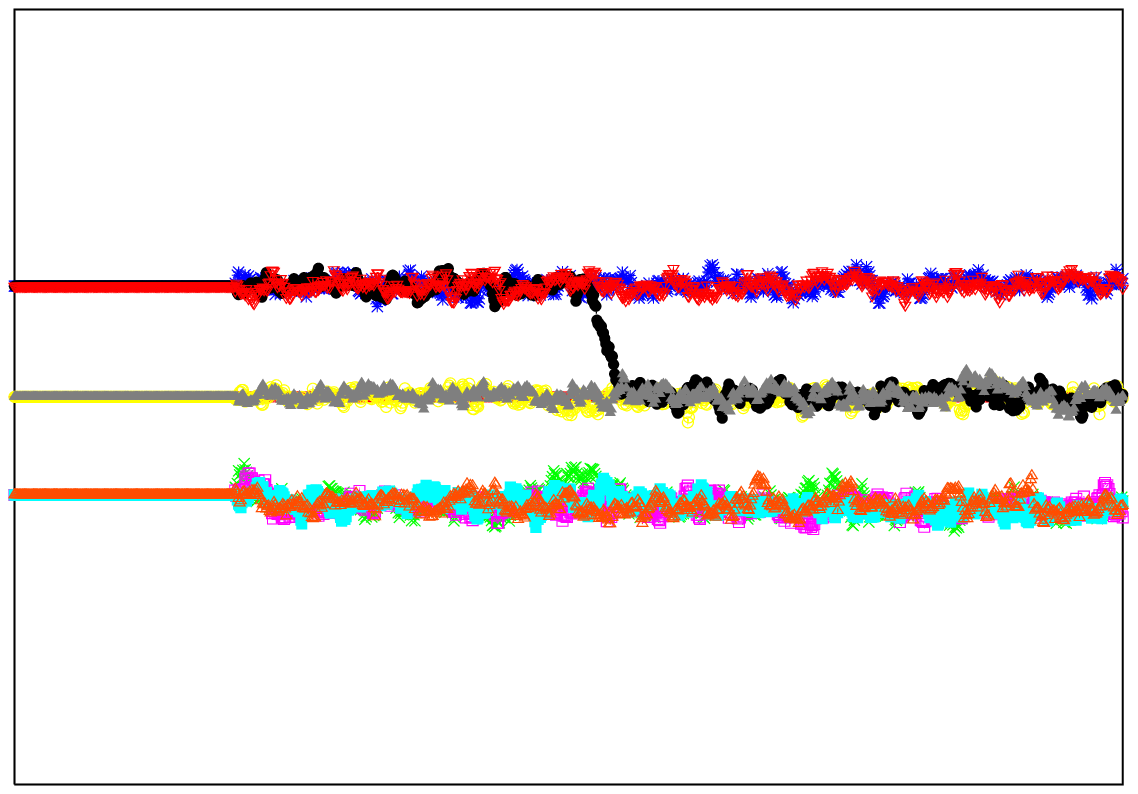}}}}
\put(0.7,3.4){\normalsize (b)}
\put(2.9,-0.2){\normalsize $u_j$}
\put(5.2,-0.2){\normalsize $1$}
\put(0.4,-0.2){\normalsize $-1$}
\put(-0.1,2.1){\normalsize $t$}
\put(0.1,0.2){\normalsize $0$}
\put(-0.2,3.6){\normalsize $100$}
\put(0.5,0.7){\normalsize $\rightarrow$}
\put(0.1,0.7){\normalsize $t^*$}
\end{picture}
\caption{\label{fig:noisy}
Typical evolution in noisy case.
(a) Noise is added from $t^*$. 
(b) Evolution of randomly selected elements in (a). 
}
\end{figure}

\subsection{Noisy case}
\begin{figure}
\begin{picture}(6,5)(0,0)
\put(0.5,0.5){\rotatebox{90}{\includegraphics[height=5.5cm,width=4cm]{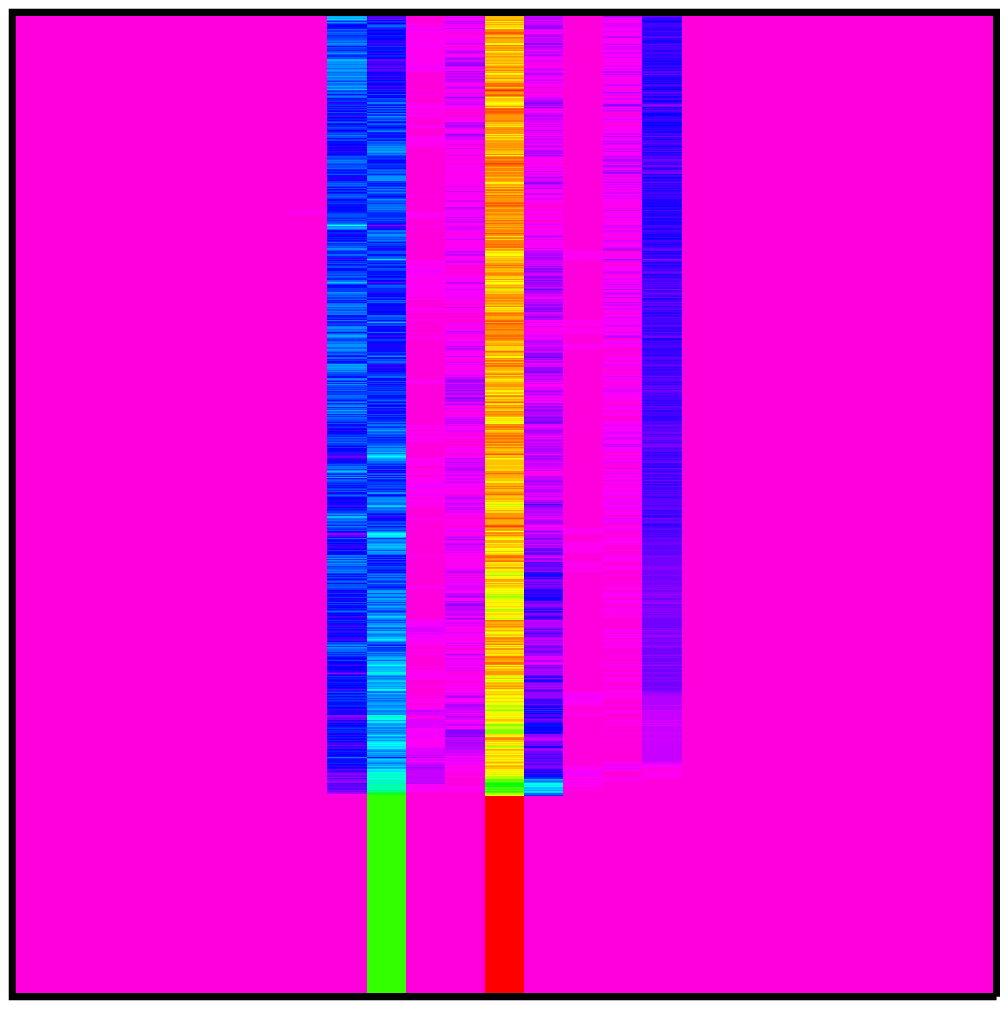}}}
\put(2.9,0.1){\normalsize $u_j$}
\put(0.4,0.1){\normalsize $-1$}
\put(5.2,0.1){\normalsize $1$}
\put(-0.1,2.2){\normalsize $t$}
\put(0.1,0.5){\normalsize $0$}
\put(-0.2,4.2){\normalsize $100$}
\put(0.5,1.2){\normalsize $\rightarrow$}
\put(0.1,1.2){\normalsize $t^*$}
\end{picture}
\caption{\label{fig:recovery}
Recovery from the invalid case shown in Fig.\ref{fig:noiseless2}(c). 
Noise ($\nu=4$) is added from $t^*$. 
}
\end{figure}

We now consider noisy case, $\nu \ne 0$. 
The noise introduced here may have 
physical, chemical or biological origin, 
like the thermal noise of the system, 
the fluctuation of distribution of the chemical substances, 
diversity of the age of the individuals
. 
Added noise kicks individuals away from the potential valley 
and escape from the local minima is expected.
Fig.\ref{fig:noisy} displays a typical 
evolution of the system in noisy case ($\nu=4$) . 
The distribution becomes broad but the ratio does not change 
(Fig.\ref{fig:noisy}(a)). 
Fig.\ref{fig:noisy}(b) shows the evolution of 
randomly selected elements as samples of behavior of individuals. 
During the time displayed, most of the elements stay in the same state. 
A few elements, however, change their internal state by the noise. 
The larger $\nu$ becomes, the more frequent 
hopping process between the states occurs. 
With too strong noise, the distribution function 
collapses into single peak.  
On the other hand, 
the system hardly escapes from the local minima 
if the noise is too weak.
In this sense there is an adequate noise amplitude.
Actually, noise added from $t^*$ set the trapped system 
free from the local minimum as shown in Fig.\ref{fig:recovery}, 
where the initial conditions are set to be trapped case 
in Fig.\ref{fig:noiseless2}(c). 

\begin{figure}
\begin{picture}(6,5)(0,0)
\put(0.5,0.5){\rotatebox{90}{\includegraphics[height=5.5cm,width=4cm]{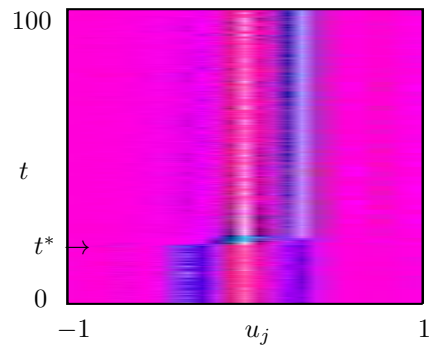}}}
\put(2.9,0.1){\normalsize $u_j$}
\put(0.4,0.1){\normalsize $-1$}
\put(5.2,0.1){\normalsize $1$}
\put(-0.1,2.2){\normalsize $t$}
\put(0.1,0.5){\normalsize $0$}
\put(-0.2,4.2){\normalsize $100$}
\put(0.5,1.2){\normalsize $\rightarrow$}
\put(0.1,1.2){\normalsize $t^*$}
\end{picture}
\caption{\label{fig:changedesign}
Ratio design is changed from 3:5:2 to 1:5:4 from $t^*$. 
Noise ($\nu=4$) are added from $t=0$.
}
\end{figure}

As another example of the adaptability of the system, 
the environment parameter is changed. 
The shape of the potential characterized by 
$(\alpha_1, \alpha_2, \alpha_3)$ 
is changed from $(0.3, 0.5, 0.2)$ to $(0.1, 0.5, 0.4)$ 
at $t^*$ (Fig.\ref{fig:changedesign}). 
The system adapts to the new environment quickly. 

\begin{figure}
\begin{picture}(6,11)(0,0)
\put(0.5,5.5){\rotatebox{90}{\includegraphics[height=5.5cm,width=4cm]{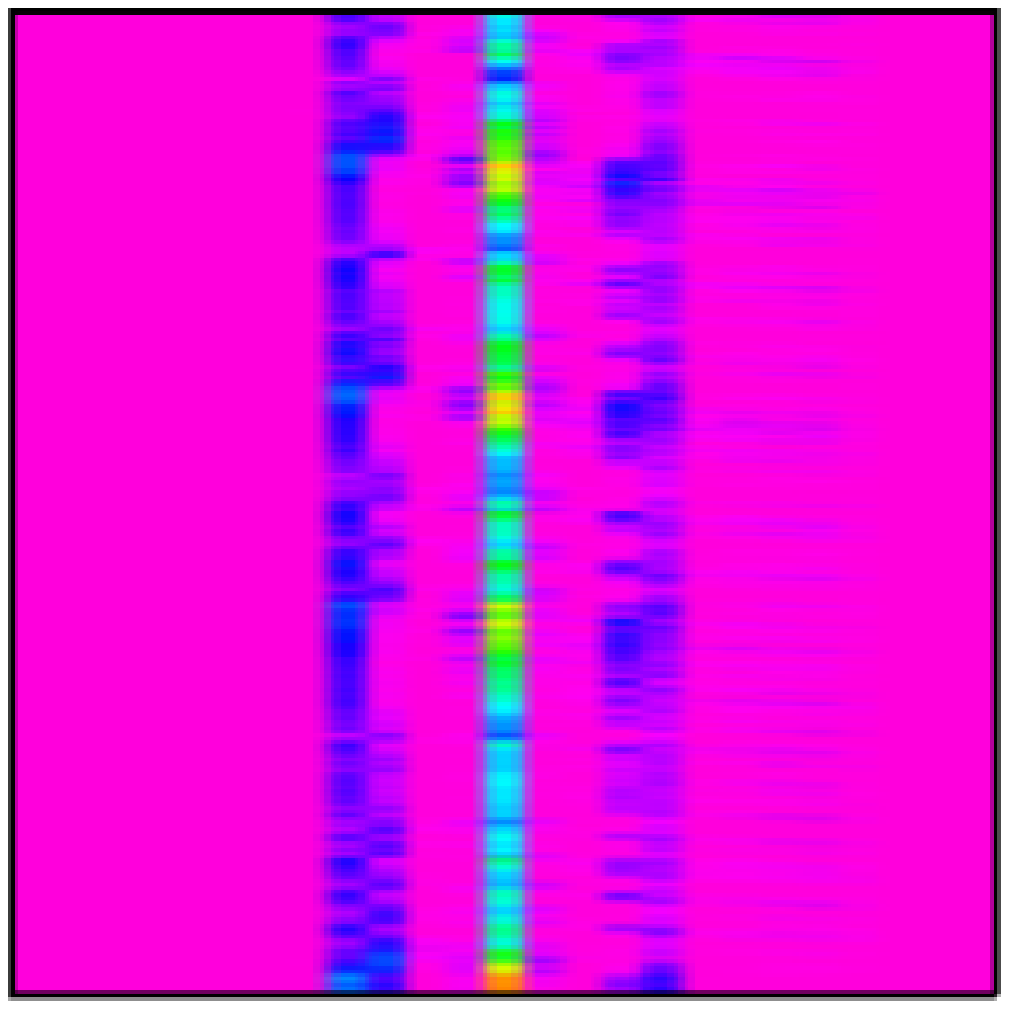}}}
\put(0.7,9.0){\normalsize (a)}
\put(2.9,5.1){\normalsize $u_j$}
\put(0.4,5.1){\normalsize $-1$}
\put(5.2,5.1){\normalsize $1$}
\put(-0.1,7.2){\normalsize $t$}
\put(0.1,5.5){\normalsize $0$}
\put(-0.4,9.2){\normalsize $1000$}
\put(1.6,10.0){\normalsize death}
\put(2.0,9.6){\normalsize $\downarrow$}
\put(4.2,10.0){\normalsize birth}
\put(4.5,9.6){\normalsize $\downarrow$}
\put(0.3,0.3){\rotatebox{90}{\includegraphics[height=5.4cm,width=4.3cm]{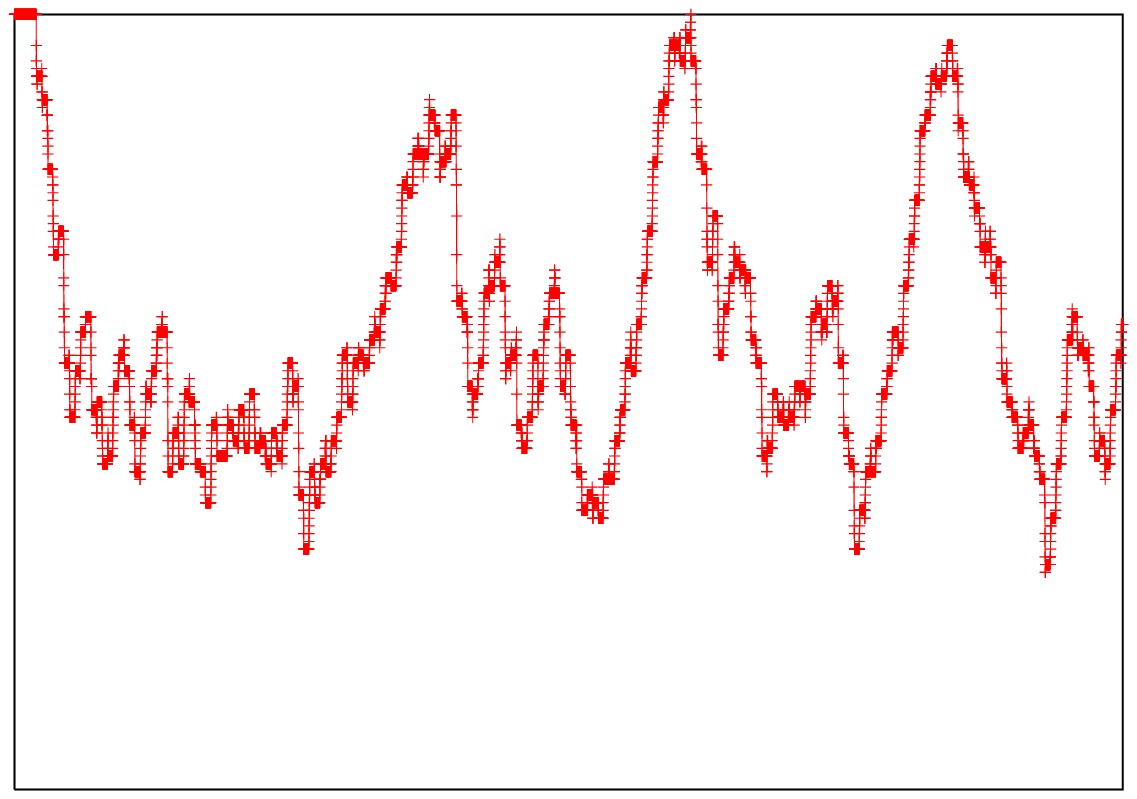}}}
\put(0.7,4.0){\normalsize (b)}
\put(2.7,0.1){\normalsize $N(t)$}
\put(5.3,0.1){\normalsize $0$}
\put(0.4,0.1){\normalsize $100$}
\put(-0.1,2.4){\normalsize $t$}
\put(0.2,0.6){\normalsize $0$}
\put(-0.4,4.2){\normalsize $1000$}
\end{picture}
\caption{\label{fig:open}
Example of ``open'' system with ``birth'' and ``death'' process.
The time series of distribution function (a) and 
total number of elements (b) are displayed.
}
\end{figure}

How about the disturbance of changing total number $N$ ?
We demonstrate an open system 
including ``birth'' and ``death'' process 
as a caricature of "polyethism" observed 
in colony of social insect. 
As a birth process of individuals, 
a new element enters into the system 
with constant internal value $u_{new} = 0.7$ 
for every five time units.
As a death process, elements which exceed 
the lower threshold $u_{death} = -0.31$ are removed 
from the system. 
Fig.\ref{fig:open} shows the typical evolution of 
distribution function and population of the whole. 
Both the proportion between the states and 
the total number are maintained with fluctuations.

\section{PHEROMONE TRAILING}\label{sec:introduction}
So far, various studies have been made on the collective behavior
of social insects \cite{Wil,De1,Sug1}. 
Especially, the trail formation of ants \cite{Wil,De1,Hel,Schw1,Bo,Nishinari,Tao} 
have called a wide interest as a remarkably synergetic 
behavior by presumably not highly intelligent individuals.
In this section,  setting a simple set of rules of which several ingredients 
are following previous studies \cite{De1,Hel}, we 
investigate the relation between i)the schedules of feeding which 
represent the unsteady natural environment, ii)emerging patterns of trails
which connect between a nest with food resources, and iii)the foraging efficiency 
as a group. Our final purpose is to connect the pattern formation of the trail
to the group strategy of foraging ants under an unsteadily varying environment.

\subsection{Model}

The model consists of a colony of N movable agents (we call them 'ants') 
which are situated on a 2D honeycomb lattice with periodic boundary
condition. Each of these ants is located at one site in the lattice   
and is heading to one of the six nearest sites(Fig.9) to which, the right-hand neighbor or the left-hand, she will move according to the 
dynamical rule described below.  
The task for ants is foraging, that is, finding food and carrying it back to a nest and the feeding sites are located at two corners of a regular triangle 
with edge-length $L$ and the nest as the third corner (fig.10).
The amount and the schedule of feeding are 
controlled through a pair of control parameters.
{$T$,   $M$} as explained in  Fig.10. 
\begin{figure}
\includegraphics[width=0.8\linewidth]{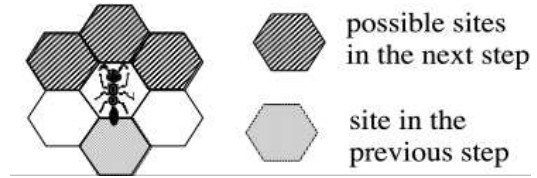}
\caption{
 Each ant is, at each time step, 
heading one of the six nearest sites.
This heading direction corresponds to the moving direction in the previous
time step. The possible moving directions in each time step are:
the forward (=facing) direction and its neighboring (right and left)
directions. To choose one of them, a stochastic rule with the weight,
$P_\alpha = exp(-\Delta^\alpha/T)/Z$
is adopted where $Z$ is the normalization factor. Here $\alpha$ indicates forward, right, or left direction  
relative to the heading direction, and
$\Delta^\alpha \equiv \rho^\alpha_\beta-\rho_\beta({\bf x},t)$ 
is the gradients of pheromone density to the candidate directions  
where $\beta$ is the index to indicate the
recruit pheromone(for mode-II ants) or the foot pheromone(for mode-III ants).
Note the above weight is not applicable for ants in mode-I
which is making random walk independent of pheromone field.
}
\end{figure}

As the mean for the communication among ants, two types of pheromones 
are secreted and perceived by individual 
ants according to their temporal modes which consist of: i)the semi-random walk mode (mode-I), ii)the exploring mode (mode-II) and 
iii)the homing mode (mode-III) (Fig.11).
The kind of pheromones considered in the present model consists of: 
i)recruit pheromone and ii)foot pheromone.
In addition these pheromones evaporate/decay in a constant rate,  also 
diffuse to the nearest sites. 

\subsection{Simulation}
All the simulations are performed on a $150 \times 150$  honeycomb lattice 
with periodic boundary condition. 
At the initial time step, N=500 ants are located at the nest, 
then simultaneously released from the nest. Note the unit time 
in our simulation corresponds to one Monte Calro step within which 
N randomly chosen ants will sequentially take one individual step. 
Under a proper combination of several fixed parameters, e.g., the evaporation/diffusion  rate of pheromones, 
and varying a pair of control parameters $\{M,T\}$, 
\begin{figure}
\includegraphics[width=0.5\linewidth]{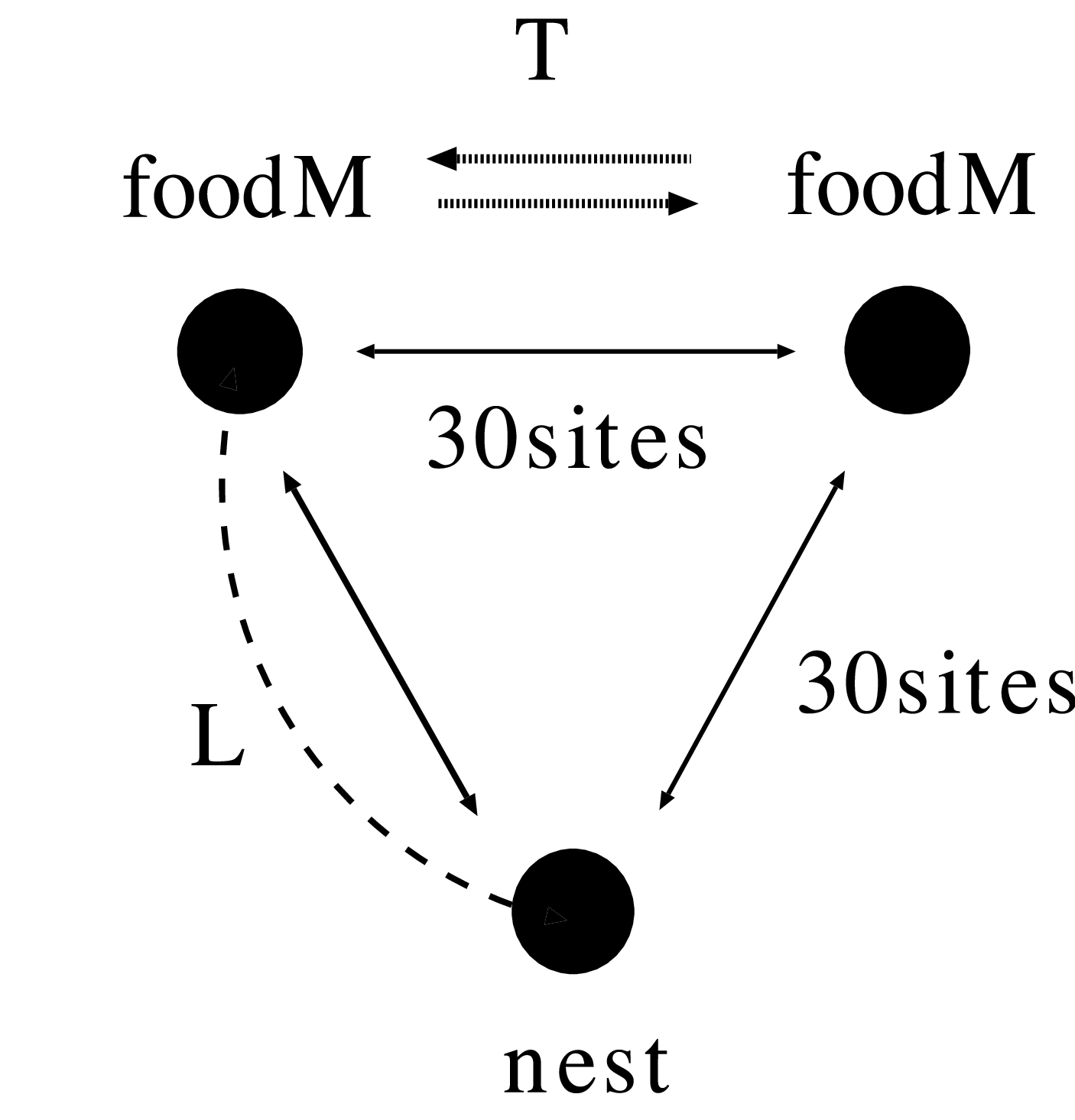}
\caption{The method of feeding. 
Feeding sites are located at two corners of an
equilateral triangle, the 3rd corner of which is the nest site.
The amount $M$ of food is supplied, 
by turn, from alternative feeding sites at every feeding interval $T$.} 
\end{figure}
\begin{figure}
\includegraphics[width=0.8\linewidth]{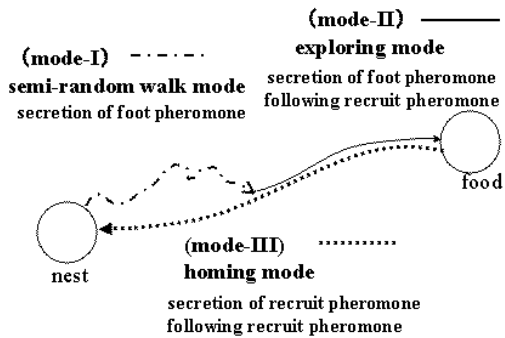}
\caption{Three modes of ants. In mode-I  ants make a semi-random walk secreting foot pheromone, 
in mode-II ants explore new food  following recruit pheromone in which mode they secrete foot pheromone, 
 in mode-III ant try to come back to nest following  foot pheromone while secreting recruit pheromone. 
The mode switch from mode-I to mode-II takes place if an ant reaches a site where the density of recruit pheromone 
exceeds a threshold value,  whereas a mode-II/mode-III  ant turns into mode-III/mode-I  on arriving  at food/nest.  
 }
\end{figure} 
the formation process of trail patterns and 
the subsequent efficient foraging are investigated.
We are particularly interested in the relation between i)the geometries of trails 
and ii)the accompanying foraging efficiency.

\bigskip
\subsection{Typical trail shapes}

Once a certain shape of trail is established it is recognized as one of following three types  
i)V-shaped trail, ii)Y-shaped trail
or iii)/-shaped trail as shown in Fig.12.
\begin{figure}
\includegraphics[width=0.7\linewidth]{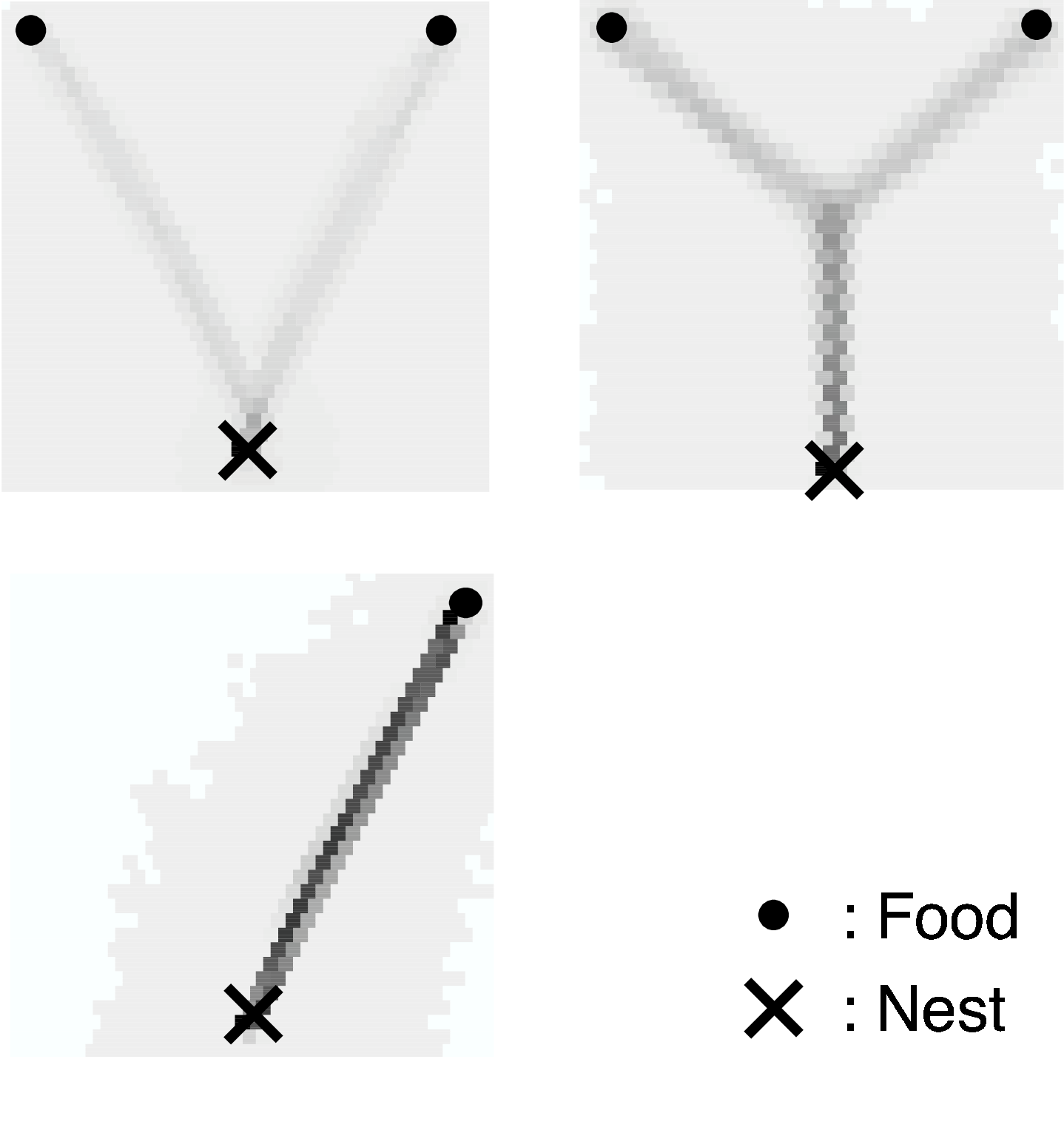}
\caption{Typical shapes of trails obtained in the simulation.
i)V-shaped trail, ii)Y-shaped trail, iii)/-shaped trail.
Here dark shading in each figure 
means the density of the ants averaged over a time significantly longer than $T$, the interval between 
successive feeding events.} 
\end{figure}
Notice that these shapes are recognizable and definable after averaging the density field of ants over a period much longer
than the feeding period $T$.  
The condition, in the $M-T$ space, for the appearance of each of these trails are shown in Fig.13 which indicates
that the most relevant quantity for the trail geometry seems to be  $M/T$. 
As $M/T$ increases, the shape of trail 
varies roughly from the V-shape to the Y-shape and finally the /-shaped
trail is obtained. 

\begin{figure}
\includegraphics[width=0.7\linewidth]{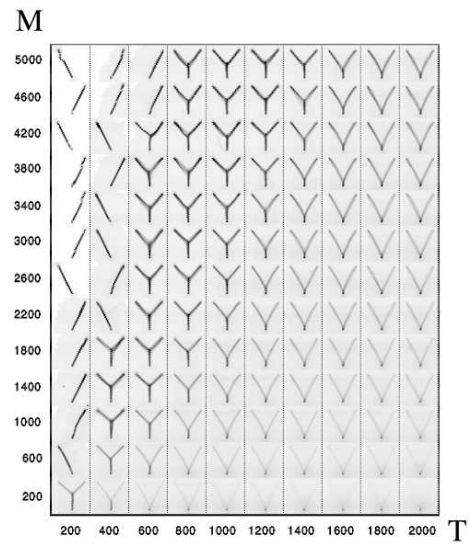}
\bigskip
\caption{The relation between emergent trail patterns 
and the combination of feeding schedule parameters,
\{$M$,$T$\}. The dark shading in each figure 
means the density of ants averaged over a time significantly longer 
than $T$ and over the ensemble of 5 simulations for the V-shaped and the Y-shaped trails.
For the /-shaped trail the ensemble average is not taken because of their intrinsic asymmetry.} 
\end{figure}

\bigskip
\noindent
\subsection{Foraging efficiency}

To investigate the relationship between the efficiency of foraging 
and trail geometry, we introduce averaged foraging efficiency $E_{av}$ 
which is the total amount of food
carried into the nest per unit time.
Hence, the relation between $M/T$ and  $E_{av}$ 
is measured under various feeding rate $M/T$ (here $T$ is fixed at $T=600$).  In figure 14, 
a characteristic relation between $M/T$ and $E_{av}$ is seen 
until $E_{av}$ reaches an saturation value.
\begin{figure}
\includegraphics[width=0.7\linewidth]{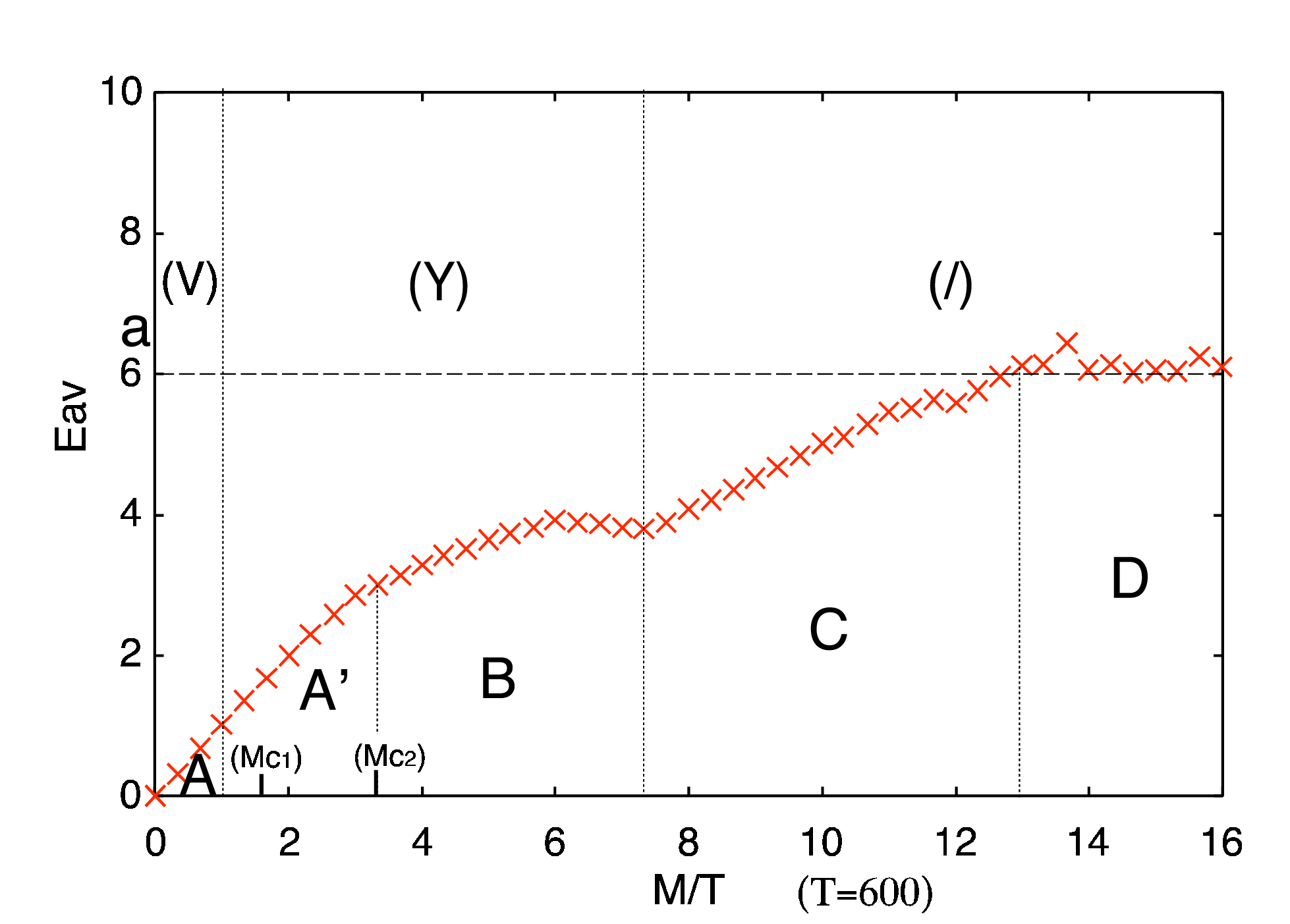}
\caption{The relation between the 
supplied amount of food per unit time $M/T$ and the long-term averaged foraging efficiency $E_{av}$.
There are five characteristic zones $A, A', B, C$ and $D$.
In zone $A$, the V-shaped trail appears, while in zones $A',B$ 
the Y-shaped trail appears, and in zones $C$ and $D$ the /-shaped trail is 
seen. Up to the zone $A'$ the {\it perfect foraging} is 
attained whereas in zone $D$ the amount of supplied food exceeds the 
carrying capacity of ants thus $E_{av}$ reaches a saturation value $a$.} 
\end{figure}
In this relation, 
considering the emergent geometry of trails, roughly five characteristic zones 
are recognized as indicated by symbols $A, A', B, C$ and $D$.
In zone $A$ with the emergence of the V-shaped trail, an almost linear $\frac{M}{T}-E_{av}$ relation is seen,   
and its slope is 1 which corresponds to the {\it perfect foraging} , namely,
ants collect all the supplied food at both feeding sites 
along the V-shaped trail.
On the other side, in zone $D$ in Fig.14, $\frac{M}{T} - E_{av}$ relation reaches a plateau, where 
the averaged foraging efficiency $E_{av}$ amounts to the maximum possible foraging efficiency $E_{max}$
realized by the present number of ants.
The last situation occurs when the amount of supplied food at each feeding site exceeds (or is equal to) the carrying capacity of 
ants, i.e., $M/2T \ge E_{max}$ is satisfied where the equivalence holds at the left edge of zone $D$.
These two zones $A$ and $D$ are characterized by simple foraging strategies with 
a straightly extending trail(a pair of straight trails) for the /-shaped trail (for the V-shaped trail).

In the case of  intermediate amount of 
food between zone A and D,  simple foraging tactics by use of straight trail(s) 
fails to give an optimal foraging.
For example, as $M$ increases over a critical value $M_{C1}$ in zone $A'$, 
if ants stick to the above mentioned  simple strategy with the V-shaped trail (a pair of straight trails), 
the sum of consuming periods for i) finding new food at alternative feeding site and  that for 
ii)carrying  away all the food at  the feeding site,  exceeds the feeding interval $T$.
It means a part of food remains to be unexhausted  even at the end of each feeding interval $T$.
However,  in our simulation at zone $A'$,  instead of sticking to such a simple strategy with the V-shaped trail, 
ants {\it invent} Y-shaped trail which,  by continuous use of the steadily maintained trunk trail
near the nest as a common path and 
leveraging the junction of Y as the 'base camp' for the exploration of  new food,
serves the system with a shorter finding time of new food  at the alternative feeding site. 
As a result, the {\it perfect foraging} is kept realized over $M=M_{C1}$ up to $M=M_{C2}$  in Fig.14\cite{note1}.
Certainly, along the Y-shaped trail,  the distance ants need to walk from the nest to food is longer than
that along a V-shaped trail.  Therefore,  once a  trail is established until all the food at the corresponding feeding site 
is exhausted,  the  temporal foraging efficiency along the Y-shaped trail falls below that along the V-shaped trail. 
In this sense, the transition from the V-shaped trail to the Y-shaped trail is a trade-off strategy within a restricted 
food environment\cite{note1}.  
This {\it perfect foraging} of both feeding sites of food is kept maintained before  a certain value 
of $M(=M_{c2}>M_{c1})$ over which no tactics can realize the perfect foraging.

In figure 15 we show the theoretically estimated 
foraging efficiencies  realized by three types of trails \cite{Tao} as the functions of $M/T$.
The figure indicates that the most efficient trail pattern varies, as $M$ increases, from V( or Y) to Y
and finally to /.  which estimation, at least qualitatively, corresponds to the simulation outcomes
shown in Fig.14. 
It is remarkable that, even in the {\it imperfect foraging regime} over $M=M_{c2}$,  through the proper change of trail geometry, 
ants in the simulation keep making the almost optimal foraging close to the theoretically estimated 
one in Fig.15 under respective feeding conditions. 
As the extreme case in zone-D, the /-shaped trail is built which is optimized in the sense that 
food is sufficient not to run out even when ants concentrate on one feeding site and in such a case 
straight trail is most efficient because of the minimal path length between the nest and food. 
Although the setup of the present simulation is very simple,  similar arguments to the present are considered to hold in more complex and 
more realistic situations like the trail formation under a larger number of feeding sites(Fig.16),  
in which situations the presumably 'unintelligent' ants are guessed to make globally efficient behavior only via local information  
carried by pheromones.   

Still,  the basic question; why such an almost optimized foraging is realized only with the pheromone 
trailing tactics,  is left unsolved.  
However it is strongly expected  that the further exploration of the pheromone trailing will serve
us with some intrinsic insights on the reason why group behavior of presumably not intellectual elements 
is kept realized in the nature.  Also, the application of the pheromone trailing tactics to 
a wider range of systems,  e.g.,  interacting multi-robot system,  is one of intriguing  issue as 
described in the next section.

\bigskip
\bigskip
\begin{figure}
\includegraphics[width=0.7\linewidth]{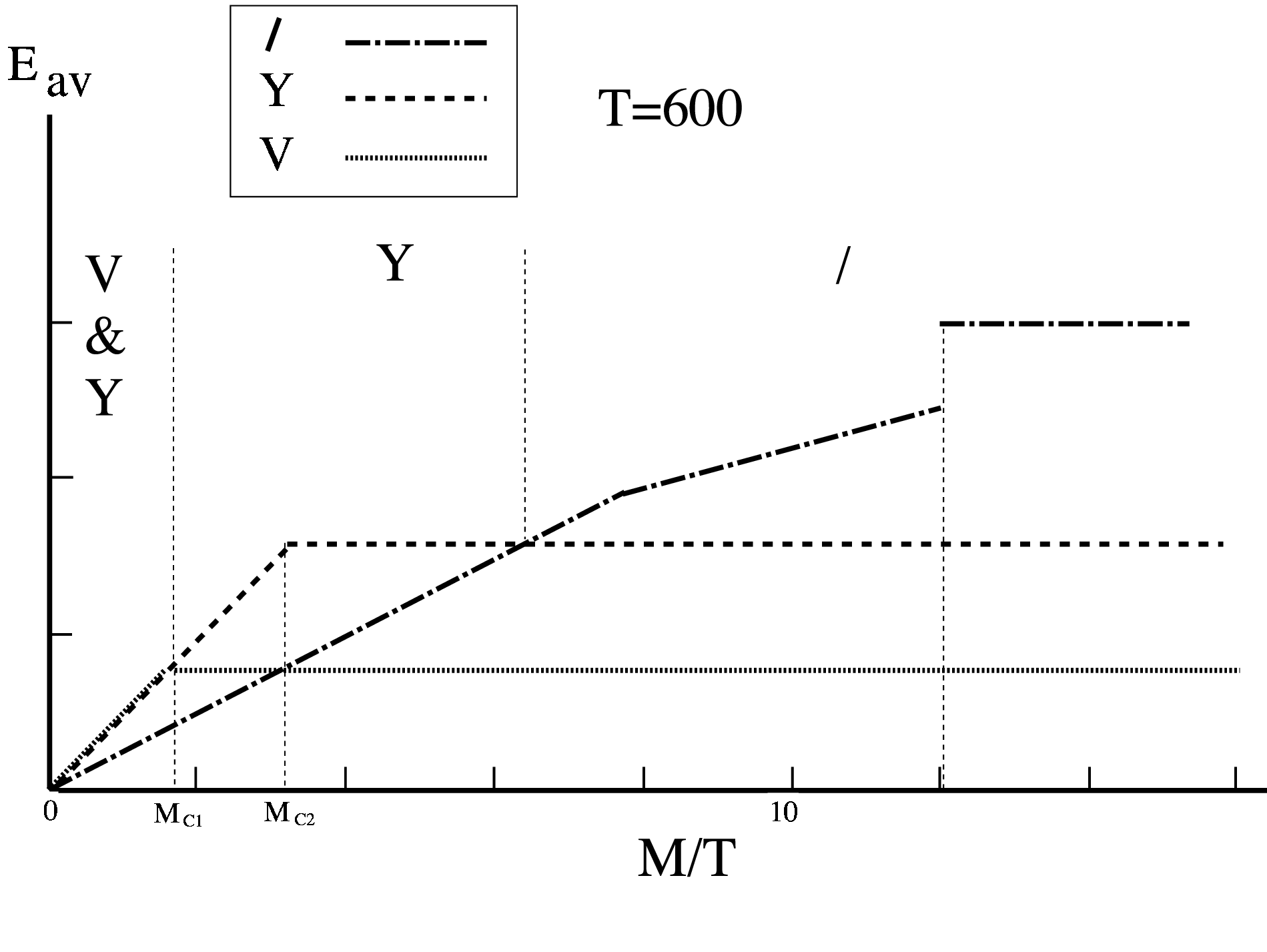}
\caption{Theoretically estimated relations between $M/T$ and averaged foraging 
efficiency $E_{av}$ for the V-shaped trail with 
the real line, for the Y-shaped trail with the dotted line,
and for the /-shaped trail with the broken line, respectively.
According to these relations, the most efficient shape of trail varies, 
as $M/T$ increases, from 
V(or Y) to Y.  Finally /-shape is seen most efficient. This way of shape shifting corresponds to the emergent trail 
shapes in our simulation with $T$=600.
Note this is the output of a crude theoretical approximation, so it contains qualitative deviations
compared to those obtained in the simulation shown in Fig.14. } 
\end{figure}
\begin{figure}
\includegraphics[width=0.5\linewidth]{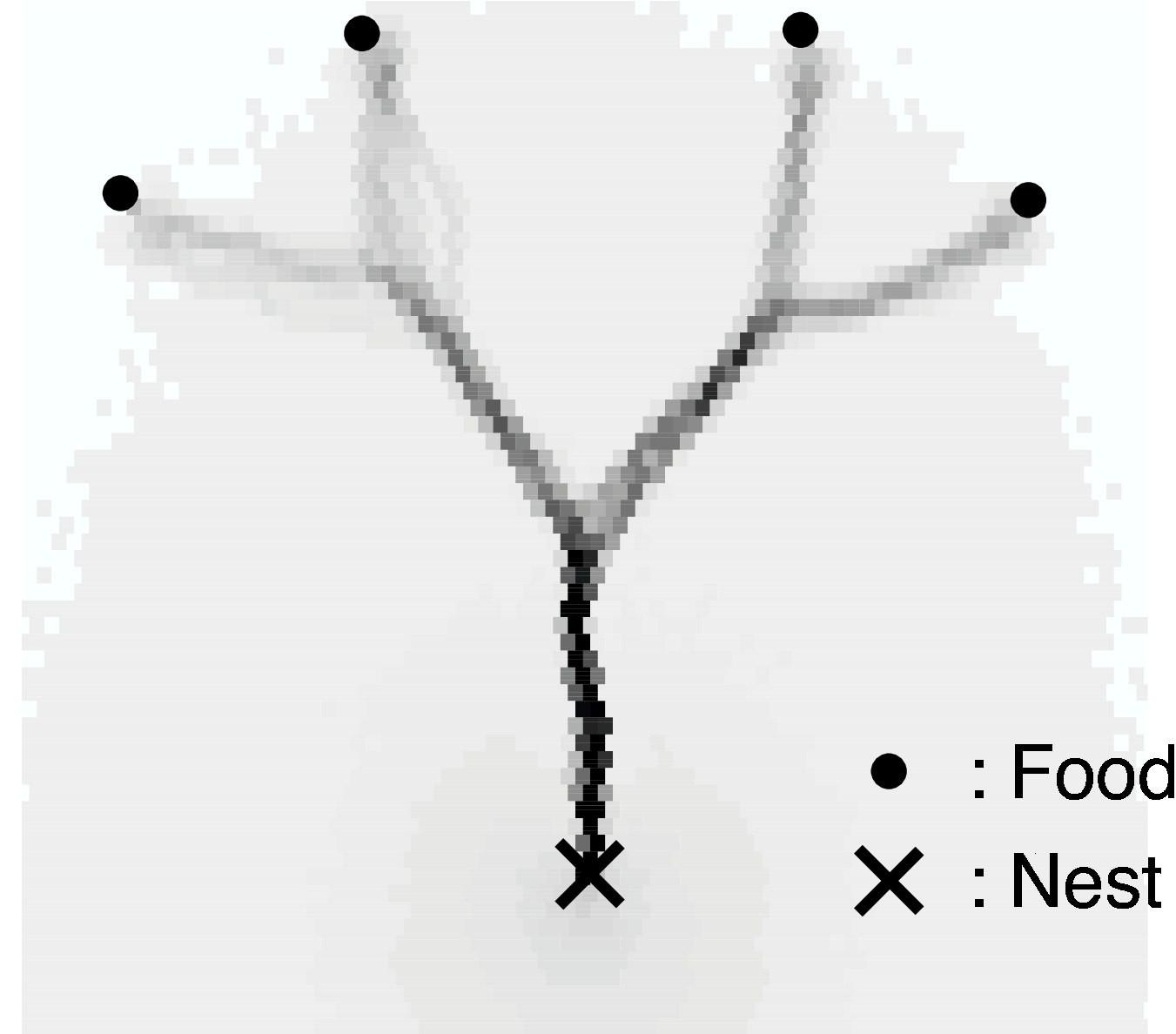}
\caption{Typical trail pattern in the case of four feeding sites}
\end{figure}

\bigskip

\section{Robot Experiments}

In previous section, we discuss the mathematical model of ant-like system, but we are also trying to develop the robots system towards the realization of ant-inspired cooperative multi-robot system. 

In engineering field, research in the area of multi-robot systems has been very active, and many researchers and engineers are currently studying the behavior of robots inspired by ant colonies\cite{Cao}. 
In most cases, physical media such as light, sound, radio wave are used for the communication between the robots, and some efforts have also been made for realizing chemical communication robot system\cite{Hayes}. 
Unfortunately, it is not easy to get proper chemical sensors at this stage. 
Moreover, chemical materials, especially gas, are invisible and it is quite difficult to observe how they spread and affect robots' behaviors. 

In our previous work, we propose "Virtual Dynamic Environment for Autonomous Robots (V-DEAR)" for real robot experiment\cite{DARS7}, in which the pheromones are virtually expressed by light information. 

\subsection{Experimental apparatus}

In V-DEAR system, pheromones are replaced with computer graphics projected on the floor.  
Robots decide their actions following the color information of the projected CG using color sensors. 
As virtual pheromones are represented as CG, we can avoid the problems described above. 
In addition, we can easily control the rate of diffusion, evaporation, diversity, etc. of the virtual chemical materials. 

\begin{figure}[h]
\includegraphics[width=\columnwidth]{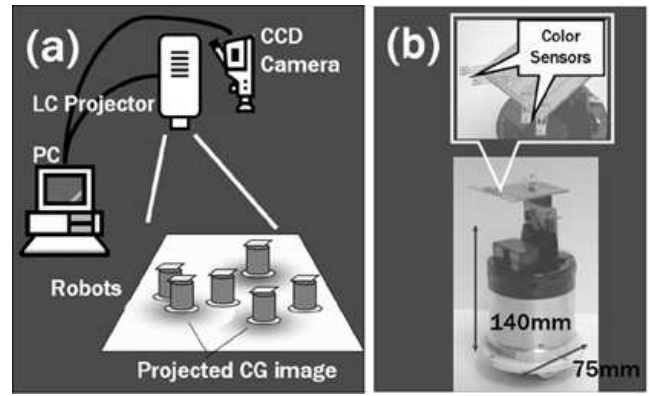}
\caption{Virtual Dynamic Environment for Autonomous Robots (V-DEAR). (a)Schematic of the system. (b)Color sensors on the top.}
\label{VDEAR}
\end{figure}

Fig. \ref{VDEAR}(a) shows the schematic of this system.
This is composed of LC projector to project the CG and the CCD camera to trace the position of the robots in the field. 
The robot moving on the field has sensors on the top to detect the color and brightness of the field(fig. \ref{VDEAR}(b)), and determines its behavior. 

Combining the position information of the robots acquired from CCD camera and the projected CG by projector, we can realize the dynamic interaction between the environment and robots.

\subsection{Task allocation}

Using this system, we made experiments of task allocation. 
In this experiment, each robot can take two states, staying and foraging, and decide one of the two states based on the total amount of food stock in the nest.
The stock is consumed constantly depending on the number of staying robots in the nest, and it is added when the foraging robot returns to the nest. 

\begin{figure}[h]
\begin{center}
\includegraphics[width=\columnwidth]{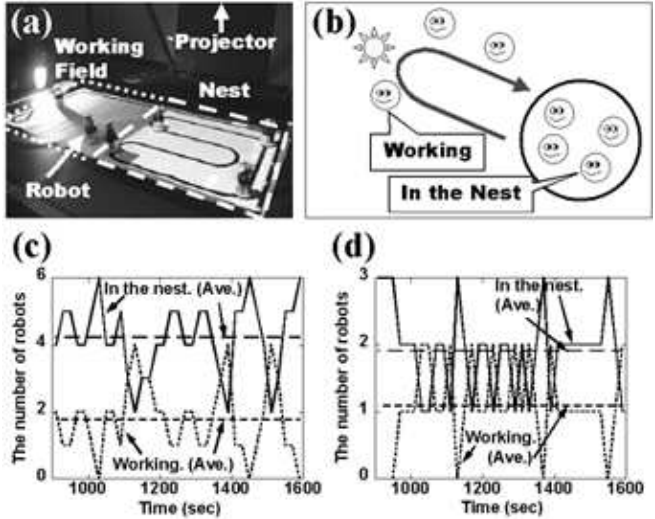}
\caption{(a)Experimental field for task allocation. (b)Schematic of the
experiment. (c)The number of robots in each state(six robots). (d)The
number of robots in each state(three robots).}
\label{fig2}
\end{center}
\end{figure}

Fig. \ref{fig2}(a) and (b) show the experimental field and the schematic, respectively. 
A half of the field is "the nest" and the robots usually move around in this area. 
The other half is "the work space" and the robots moving around this area carry back food. 
V-DEAR system monitors the total amount of food stock and expresses it by the brightness of the nest. 
Each robot detects it and decides their state autonomously. 
The robots were adjusted to show nearly 1/3 working ratio in case of six robots system (Fig. \ref{fig2}(c)). 
We reduced the number of robots by half under the same condition, but you can see the rate is kept 1/3 approximately (Fig. \ref{fig2}(d)).

\subsection{Pheromone Trailing}

\begin{figure}[h]
\begin{center}
\includegraphics[width=\columnwidth]{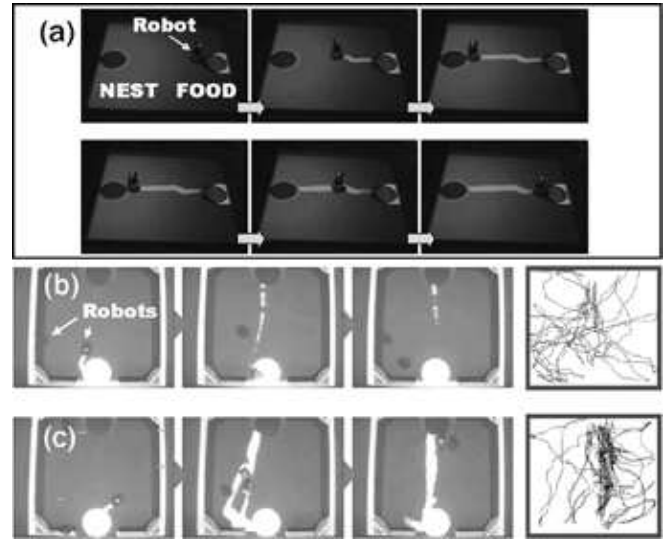}
\caption{(a)Basic behavior of the foraging robot. On discovering food, the robot lays chemical trail while returning to nest. The robot detecting the pheromone follows the trail.
(b)In case that the rate of the evaporation is high. 
(c)In case that the rate of the evaporation is low.}
\label{fig3}
\end{center}
\end{figure}

We also made foraging experiments by multi-robot system. 
Fig. \ref{fig3}(a) shows the basic behavior of the foraging robots. 
On discovering a food, the robot turns on a LED on the top and moves towards the nest. 
The V-DEAR system detects the LED and projects a CG pheromone trail during the LED is turned on. 
When the robot arrives at the nest, it turns off the LED and changes into the searching state. 
If it finds the pheromone trail, it follows the trail. 

Fig. \ref{fig3}(b)(c) show the snapshot of the foraging by two robots in case of high evaporation rate and low evaporation rate, respectively, and the drawings at the right endpoint are the trajectories of the robots. 
In case that the evaporation rate of the pheromone is high, the pheromone trail hardly remains. 
However, the evaporation rate is low, a stable trail is formed between the food point and their nest.

\section{summary}
Here in this article, we have investigated two topics 
among intriguing behaviors exhibited by biological mass. 
In \S 2, we proposed a model for multi(three)-tasks 
allocation in which individuals feel 
the situation of the whole of the colony 
through the stock materials $w_j(t)$ 
and determines its behavior using 
its internal variables $u_i(t)$ and potential $U(t)$. 
The proportion regulation against several 
types of disturbances was exhibited qualitatively. 
As for general task rule like more complex ones or 
especially asymmetric ones, 
further improvement seems to be required.

In \S 3, foraging  behavior of ants was simulated, in which 
ants rely only on the local information of two kinds of pheromones and 
move according to a set of primitive  rules. 
However, they fulfill a highly efficient foraging flexibly adapting to various feeding conditions.
Although present simulations are restricted to rather simple
cases,  same kind of argument is expected to be available for
more complex and more realistic situations.

In \S 4, small scale experiments of 
task allocation process and trail formation process 
were examined using multi-robot and V-DEAR system 
which simulates pheromones by light information.

The combination of the above investigations
would uncover the fundamental mechanism for
the formation, the maintenance and the evolution
of various types of collective dynamics exhibited by
a wide class of active elements.

\begin{acknowledgements}
This work is supported by the 
Japanese Grand-in-Aid for Encouragement of Young Scientists from the
Ministry of Education, Science and Culture (No.15760291 and No.16605008) 
\end{acknowledgements}



\begin{thebibliography}{99}
\bibitem{Raper}K.B. Raper and J.Elisha, {\it Mitchell Scient. Soc.} {\bf 56} (1940) 241.
\bibitem{Bonner}J. T. Bonner, {\it Q. Rev. Biol.} {\bf 32} (1957) 232; {\it The Cellular Slime Molds}, Princeton Univ. Press, Princeton, New Jersey, 1967, 2nd ed.
\bibitem{Loomis}W. F. Loomis, {\it Dictyostelium discoideum: A developmental System}, Academic Press, New York, 1975.
\bibitem{Oyama} M. Oyama, K. Okamoto and I. Takeuchi, {\it J. Embryol. Exp. Morph. }{\bf 75} (1983) 293.
\bibitem{Wilson2}E.O. Wilson, {\it Sociobiology}, Harvard, Cambridge, MA, 1975.
\bibitem{Edelstein-Keshet}L. Edelstein-Keshet, 
in: W. Alt (Ed.), {\it Lecture Notes in Biomathematics, Vol.89}, Springer, Berlin, 1990, p.528.
\bibitem{Partridge}B. L. Partridge, {\it Sci. Am.} {\bf 246} (1982) 90.
\bibitem{Inoue}M. Inoue, {\it Schooling of Fishes; behavior} (Kaiyo-shuppan, Tokyo, 1981) (in Japanese).
\bibitem{Shimoyama}N. Shimoyama et al. {\it Phys. Rev. Lett.}, {\bf 76} (1996) 3870.
\bibitem{Vicsek}T. Vicsek, A. Czirok, E. Ben-Jacob, I. Cohen, and O. Shochet, {\em Phys. Rev. Lett.}, {\bf 75} (1995) 1226--1229.
\bibitem{Wilson}E.O. Wilson, {\it The Insect Societies}, Oxford Univ. Press, London, 1971.
\bibitem{GlobalTuring}T. Mizuguchi and M. Sano, {\it Phys. Rev. Lett.}, {\bf 75} (1995) 966.
\bibitem{Cao} Y.U. Cao, A.S. Fukunaga and A.B. Kahng, "Cooperative Mobile Robotics:Antecedents and Directions," {\em Autonomous Robots}, 4, (1997) pp.7-27.
\bibitem{Hayes} A.T. Hayes, A. Martinoli, and R.M. Goodman, "Distributed Odor Source Localization", {\em IEEE Sensors}, Vol. 2, No. 3 (2002) pp.260-271. 
\bibitem{DARS7} T. Kazama, K. Sugawara, and T. Watanabe, "Collecting Behavior of Interacting Robots with Virtual Pheromone," {\em Proc. 7th Int. Symp. on
Distributed Autonomous Robotic Systems}, (2004) pp.331-340.
\bibitem{Wil}
B.H\"{o}lldobler B.and Wilson E.O.
{\it The Ants}
{Belknap, Cambrige}
(1990)


\bibitem{De1} 
{Deneubourg J.L.and Goss S.and Franks N. and Pasteels J.M.} 
REVIEW{J. Insect Behavior} \textbf{2}(1989) 719.

\bibitem{Sug1}
{N.Shimoyama, K.Sugawara, T.Mizuguchi, Y.Hayakawa, and M.Sano} 
{Phys. Rev. Lett.} \textbf{76} (1996) 3870.


\bibitem{Hel} 
{D.Helbing and F.Schweitzer and P.Moln\'{a}r}
{Phys. Rev. E} \textbf{56}{1997} 2527.

\bibitem{Schw1}
{F.Schweitzer, K.Lao and F.Family} 
{BioSystems} \textbf{41} (1997) 153.

\bibitem{Bo} 
{E.Bonabeau, M.Dorigo and G.Theraulaz}
{Swarm Intelligence}
{Oxford University press, Oxford}
{1999}.


\bibitem{Nishinari}
{K.Nishinari, D.chowdhury and A.Schadschneider} 
{Phys. Rev. E} \textbf{67} (2003) 36120.


\bibitem{Tao}
{T.Tao, H.Nakagawa, M.Yamasaki and H.Nishimori} 
{J. Phys. Soc. Jpn.} \textbf{73} (2004) to be pubslished.

\bibitem{note1} 
{Note that, in our simulation,  the tranition value of $M(=M_c)$  from  the V-shaped trail to the Y-shped trail is
not just the  value $M_{C1}$ over which  the perfect foraging using the V-shaped trail gets impossible 
but lower than that. 
It dose not mean, however, the abortion of optimized foraging at $M_c < M < M_{c1}$. 
According to our theoretical analysis in ref\cite{Tao}.,  if  $M<M_{c1}$   
the foraging efficiency is equivalent between the foraging with the Y-shaped trail and that with V-shaped trail 
as shown in Fig.14.}
\end{thebibliography}
\end{document}